\documentclass{iaus}
\usepackage{graphicx}

\title[Spitzer observations of deeply obscured galactic nuclei]
{Spitzer observations of deeply obscured galactic nuclei}

\author[Spoon et al.]
{H.W.W. Spoon$^1$,
 J.V. Keane$^2$,
 J. Cami$^3$,
 F. Lahuis$^4$,\break
 A.G.G.M. Tielens$^5$,
 L. Armus$^6$, \and
 V. Charmandaris$^7$}

\affiliation{
$^1$Cornell University, Astronomy Department, Ithaca, NY 14853, USA 
\break email: spoon@astro.cornell.edu \\[\affilskip]
$^2$ NASA Ames Research Center, MS 245-3, Moffett Field, CA 94035 \\[\affilskip]
$^3$ NASA Ames Research Center, MS 245-6, Moffett Field, CA 94035 \\[\affilskip]
$^4$ Leiden Observatory, P.O. Box 9513, 2300 RA Leiden, Netherlands \\[\affilskip]
$^5$ SRON and Kapteyn Institute, P.O. Box 800, 9700 AV Groningen, The Netherlands\\[\affilskip]
$^6$ Caltech, Spitzer Science Center, MS 220-6, Pasadena, CA 91125, USA\\[\affilskip]
$^7$ Department of Physics, University of Greece, P.O. Box 2208, 71003 Heraklion, Greece}

\date{November 15 and in revised form ??}

\pubyear{2005}
\volume{231}  
\pagerange{281--290}
\jname{Astrochemistry: Recent Successes and Current Challenges}
\editors{D.C. Lis, G.A. Blake \& E. Herbst, eds.}

\begin{document}

\maketitle


\begin{abstract}
We report on our first results from a mid-infrared spectroscopic 
study of ISM features in a sample of deeply obscured ULIRG nuclei 
using the InfraRed Spectrograph (IRS) on the Spitzer Space Telescope. 
The spectra are extremely rich and complex, 
revealing absorption features of both amorphous and crystalline 
silicates, aliphatic hydrocarbons, water ice and gas phase bands 
of hot CO and warm C$_2$H$_2$, HCN and CO$_2$. PAH emission 
bands were found to be generally weak and in some cases absent.
The features are probing 
a dense and warm environment in which crystalline silicates and 
water ice are able to survive but volatile ices, 
commonly detected in Galactic dense molecular clouds, cannot.
If powered largely by star formation, the stellar density and
conditions of the gas and dust have to be extreme not to 
give rise to the commonly detected emission features associated 
with starburst.

\keywords{ISM: evolution --  ISM: molecules -- galaxies: ISM -- 
galaxies: nuclei -- infrared: ISM}

\end{abstract}


\firstsection 
\section{Introduction}

With the launch of the InfraRed Spectrograph (IRS) on the Spitzer
Space Telescope, mid-infrared spectroscopists have been handed a 
powerful tool to study extragalactic objects at high signal-to-noise 
and over a wide mid-infrared wavelength range, previously only
available for the study of Galactic sources and a few nearby 
galaxies. 

The foundations for the Spitzer studies currently underway were
laid by the Infrared Space Observatory (ISO), which freed 
mid-infrared spectroscopists from the confinements of the Earth's 
atmospheric windows and the atmosphere's glaring foreground 
emission.

Major extragalactic topics addressed early on in the ISO mission 
centered around the properties of dusty starbursts 
and how they evolve (\cite[e.g. Thornley et al. 2000]{thornley00}), 
the unification of optically classified type 1 and 2 active 
galaxies in relation to the properties of the Active Galactic 
Nucleus (AGN) (\cite[e.g. Clavel et al. 2000; 
Laurent et al. 2000]{clavel00,laurent00}), 
and the dominant power source in Ultra-Luminous Infrared Galaxies 
(ULIRGs): massive starbursts or AGN activity 
(\cite[e.g. Genzel et al. 1998; Tran et al. 2001]{genzel98,tran01})?

After the expiration of ISO, two unusual galaxy spectra
provided first mid-infrared insights into the properties of gas 
and dust in deeply obscured galactic nuclei. 
The 2--5\,$\mu$m spectrum of the nucleus of NGC\,4945 revealed 
strong absorptions of water ice (3\,$\mu$m), CO$_2$ (4.26\,$\mu$m) 
and a blend of `XCN' and CO ice (4.65\,$\mu$m), seen against a 
continuum obscuring the central massive black hole 
(\cite[Spoon et al. 2000]{spoon00}).
Groundbased follow-up observations confirmed the 4.65\,$\mu$m 
absorption band to consist of separate components of processed 
`XCN' and CO ice and warm (35\,K) CO gas. The detection of 
processed ices against the nuclear continuum is taken as an
indication for the presence of dense star forming molecular clouds 
towards the nucleus of this active galaxy (\cite[Spoon et al. 2003]{spoon03}).

The second unusual spectrum is that of the nucleus of NGC\,4418,
originally classified by \cite{roche86} as a ``very extinguished 
galaxy'', for its very deep 10\,$\mu$m silicate absorption feature. 
Instead of the commonly detected PAH emission features, the 
5.5--8\,$\mu$m ISO spectrum of its nucleus is dominated 
by absorption features
of water ice (6\,$\mu$m), hydrocarbons (6.85\,$\mu$m and 
7.25\,$\mu$m) and methane ice (7.67\,$\mu$m), reminiscent of the
line of sight towards our own Galactic Center 
(\cite[Spoon et al. 2001]{spoon01}).
Supporting groundbased observations indicate that most of the
infrared luminosity (L$_{\rm IR}$\,=\,10$^{11}$\,L$_{\odot}$) 
from NGC\,4418 is produced in a compact nucleus with a radius of 
less than 
80\,pc (\cite[Evans et al. 2003]{evans03}). If powered entirely 
by star formation, the conditions of the nuclear gas and dust 
within this environment must be exceptional not to give rise to 
emission features commonly associated with star formation.

\begin{figure}
\begin{center}
\includegraphics{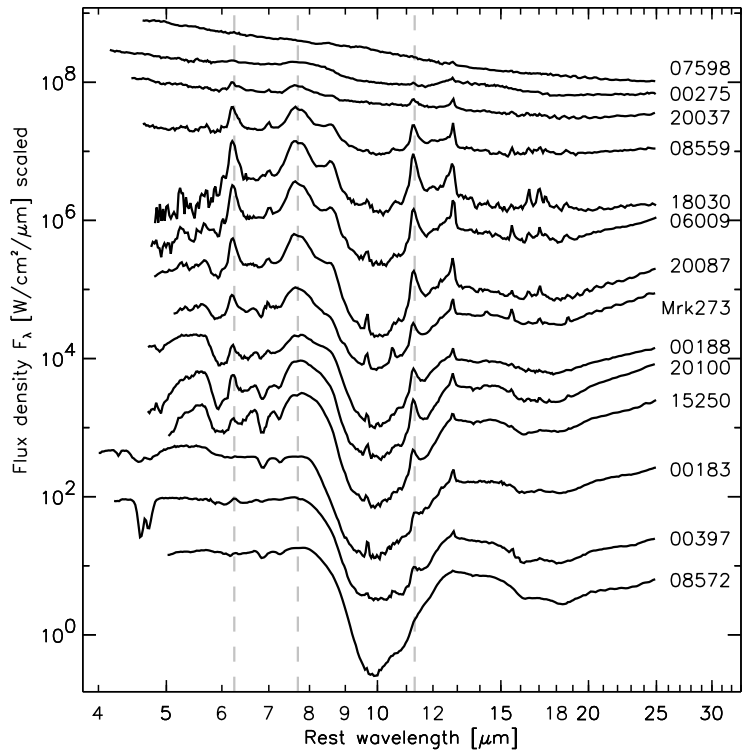}
\end{center}
\caption{Spitzer IRS low-resolution spectra of ULIRGs sorted by
spectral shape. The three spectra at the top are continuum-dominated
(AGN-like), the next four are PAH-dominated (starburst-like) and
the rest are absorption-dominated (burried nuclei). Vertical lines 
indicate the positions of the 6.2, 7.7 and 11.2\,$\mu$m PAH 
emission bands}
\label{fig:sed_trends}
\end{figure}

A large scale follow-up study into the presence of 5.5--8\,$\mu$m 
absorption features in ISO galaxy spectra resulted in the finding
of 6\,$\mu$m water ice absorption in 12 out of 19 ULIRGs, 2 out of 
62 AGNs and 4 out of 21 starburst galaxies surveyed. Also, 6.85\,$\mu$m 
hydrocarbon absorption was detected in three galaxies besides 
NGC\,4418, all three of which are ULIRGs.
The results are consistent with findings from other wavelength
ranges that more molecular material is present in ULIRG nuclei
than in other galaxy types (\cite[Spoon et al. 2002]{spoon02}).

In the following sections we present an overview of the first results 
from an IRS spectroscopic study of the properties of gas and dust 
in strongly obscured ULIRG nuclei. 
The spectra were selected from a larger sample of ULIRG spectra 
obtained as part of the GTO program of the Spitzer IRS team.

\section{The diverse nature of the ULIRG family}

Figure\,\ref{fig:sed_trends} offers a strikingly illustration of the 
diverse nature of the galaxies classified as ULIRGs. At the top of
the figure we find ULIRG mid-infrared spectra dominated
by AGN-heated hot dust, with IRAS\,07598 displaying a weak silicate 
emission feature. Further down, the spectra of IRAS\,20037 and
IRAS\,08559 show increasing contributions of PAH emission superimposed 
on the AGN continuum. Moving down to IRAS\,18030 and 
IRAS\,06009, PAH emission starts to dominate and silicate absorption 
at both 10 and 18\,$\mu$m becomes apparent. In the spectrum of
the next one down, IRAS\,20087, the characteristic absorption edge 
of water ice at 5.7\,$\mu$m starts to appear. The water ice feature 
deepens and the importance of PAH emission decreases moving down 
from Mrk\,273 to IRAS\,15250. At the same time, the depth of the 
10 and 18\,$\mu$m silicate features increases and hydrocarbons
absorption bands at 6.85 and 7.25\,$\mu$m become apparent. Further
note how from IRAS\,18030 to IRAS\,15250 the characteristic 
PAH emission feature at 7.7\,$\mu$m gradually gets replaced by a 
broad absorbed-continuum peak at 8\,$\mu$m. Starburst diagnostics
relying on the equivalent width or luminosity of the 7.7\,$\mu$m 
PAH feature will have to carefully verify the nature of this peak 
(\cite[Spoon et al. 2004a]{spoon04a}). The bottom three
spectra in figure\,\ref{fig:sed_trends} differ from those directly
above by the presence of a strong near-infrared continuum and by
the relative weakness of the 5.5--8\,$\mu$m absorption features. 
Note how, due to the appreciable redshifts of IRAS\,00183 and 
IRAS\,00397, the IRS spectral coverage extends all the way down 
to rest frame 4\,$\mu$m, facilitating the discovery of wide 
absorption features of CO gas in their spectra 
(\cite[Spoon et al. 2004b]{spoon04b}).
Recent groundbased high-resolution M-band spectroscopy reveals the
CO band to be also present in the spectrum of the low-redshift ULIRG
IRAS\,08572 (Geballe et al., in preparation).

\section{Absorption features of crystalline silicates}

\begin{figure}
\begin{center}
\includegraphics[width=10.5cm]{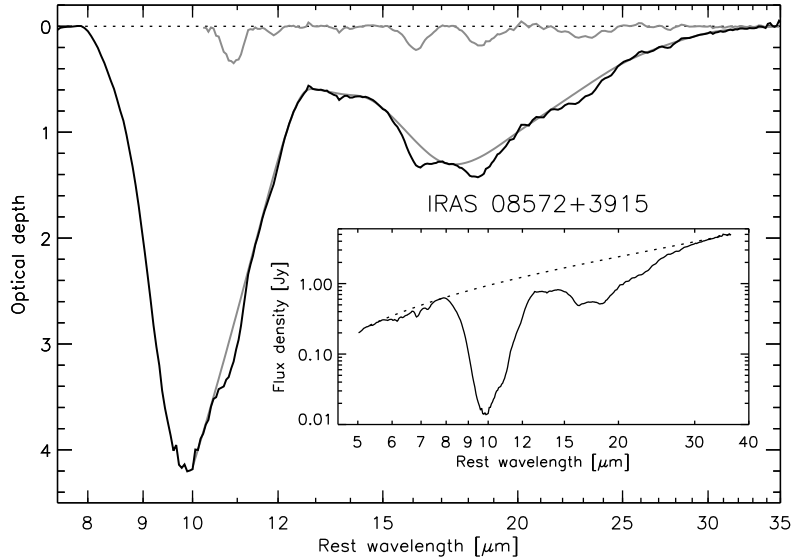}
\end{center}
\caption{Silicate optical depth spectrum of IRAS\,08572 ({\it black})
with a spline fit to the amorphous silicate component overplotted
({\it grey}). The residual spectrum, revealing crystalline features
at 11, 16, 19 and 23\,$\mu$m is shown in {\it grey} at the top 
of the plot. Inset: the 5--35\,$\mu$m spectrum of IRAS\,08572 
with the adopted local continuum.}
\label{fig:08572_panel}
\end{figure}

\begin{figure}
\begin{center}
\includegraphics[width=11cm]{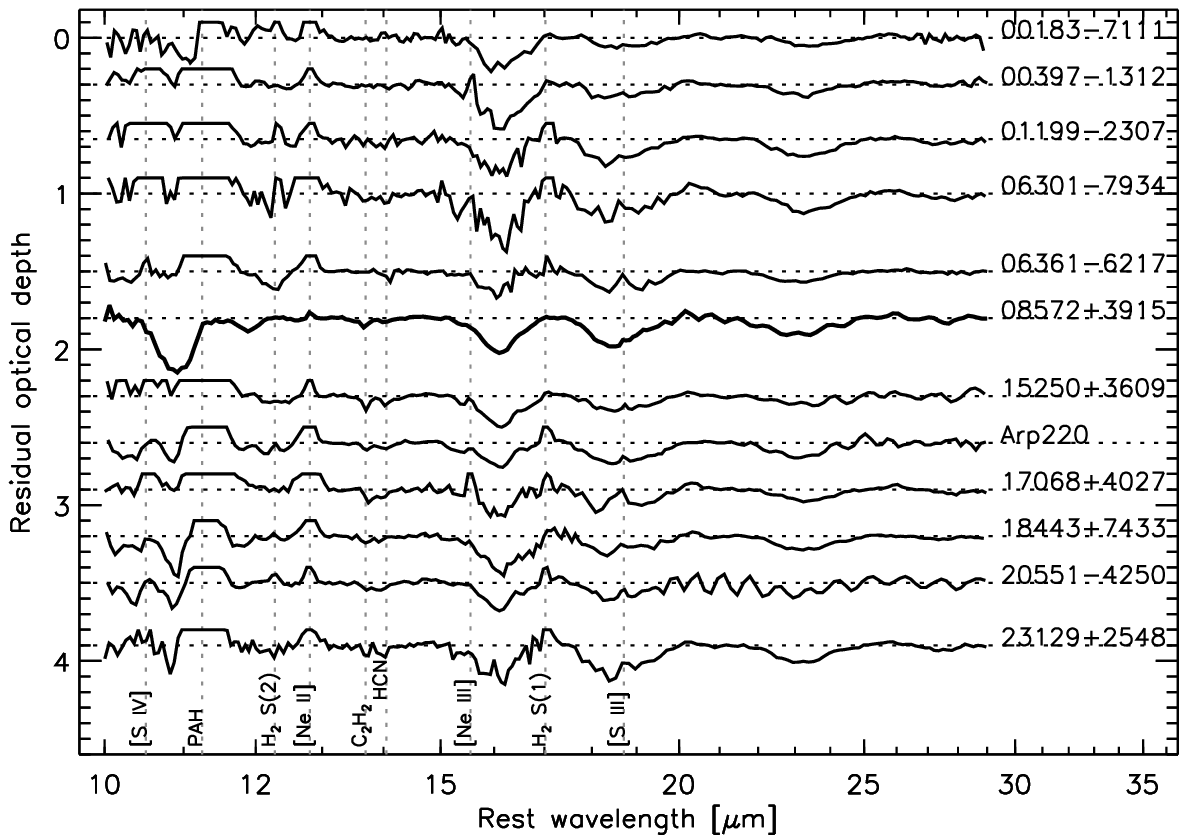}
\end{center}
\caption{10--30\,$\mu$m residual optical depth spectra for 12
ULIRGs from our sample after subtraction of the amorphous silicate 
component. For plotting purposes,
the spectra have been offset and truncated at residual optical 
depths of -0.1. Crystalline silicate features can be identified 
at 11, 16, 19, 23 and 28\,$\mu$m}
\label{fig:crystsil_coll}
\end{figure}

We have detected significant substructure in the deep silicate absorption 
features towards a sample of strongly obscured ULIRG nuclei. The absorption
features appear at 11, 16, 19, 23 and 28\,$\mu$m and are best revealed
by a spline fit to the amorphous silicate absorption component. 
We identify the residuals with crystalline silicates, most likely 
forsterite (Mg$_2$SiO$_4$). 
The clearest detection is offered by the spectrum of IRAS\,08572, shown 
in detail in figure\,\ref{fig:08572_panel}. This spectrum combines the
deepest silicate feature in our sample with the complete absence of all
commonly present PAH emission features. In figure\,\ref{fig:crystsil_coll}
we show our 12 clearest crystalline silicate detections. For these we
infer crystalline to amorphous silicate ratios of 7--15\%, using the
optical depth and opacities of the 10\,$\mu$m amorphous and 16\,$\mu$m
crystalline band.

The detection of crystalline silicates appears to be geared 
to the more deeply enshrouded ULIRG spectra. While all 17 ULIRGs with
$\tau$(sil)\,$>$\,2.9 do show the strong 16\,$\mu$m feature, at lower 
silicate optical depth the number of detections of this band drops 
sharply. Between $\tau$(sil)=2.0 and $\tau$(sil)=2.9 only 1 out of 6
ULIRGs shows the feature and below that, only 3 out of 54. This also
is apparent in the spectra shown in figure\,\ref{fig:sed_trends}, 
in which the 16\,$\mu$m feature can only be recognized in the spectra 
with the deepest silicate features. 

The discovery of crystalline silicates in the ISM of ULIRGs stands 
in sharp contrast to the upper limit of 1\% for the crystallinity
of the Galactic ISM (\cite[Kemper et al. 2004, 2005]{kemper04,kemper05}).
As stellar sources of silicate dust inject at least 5\% of their 
silicates in crystalline form, the crystalline 
silicates arriving in the Galactic ISM must be rapidly transformed into 
amorphous silicates. The timescale for amorphization is estimated 
to be only 70 million years (Bringa et al., in preparation), considerably 
shorter than the timescale at which (crystalline) silicates are 
injected into the Galactic ISM (4 billion years; Bringa et al., 
in preparation).
In ULIRGs and other mergers these time scales may be different. 
ULIRGs are characterized by a high rate of star formation driven 
by merging events. In contrast to the Galactic ISM, the
enrichment of the dusty interstellar medium will be dominated by the
rapidly evolving massive stars and the contribution by the more
numerous low-mass stars will lag on the timescale associated with the
ULIRG-merger event (\cite[$10^8-10^9$ yr; Murphy et al. 1996]{murphy96}). 
Thus, we attribute the high fraction 
of crystalline silicates in these ULIRGs as compared to others 
to the relative `youth' of these systems; e.g., the amorphitization
process may lag the merger-triggered, star-formation-driven dust 
injection process.
For a more detailed discussion on the discovery of crystalline
silicates in ULIRG spectra we refer to \cite{spoon06}.

\section{Absorption features in the 5.5--8\,$\mu$m range}

\begin{figure}
\begin{center}
\includegraphics{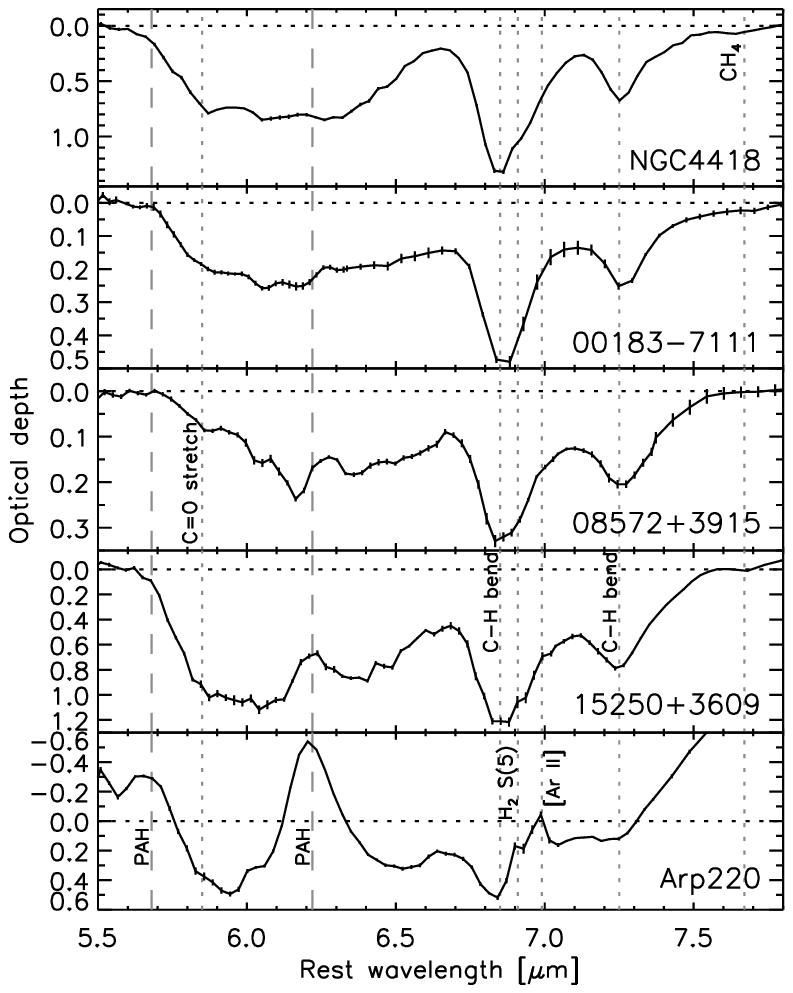}
\end{center}
\caption{Optical depth spectra for five (U)LIRGs of absorption
features in the 5.5--7.7\,$\mu$m range. Vertical {\it dashed} lines
denote the central wavelengths of PAH emission features at 5.7 and
6.22.
Vertical {\it dotted} lines indicate the central positions of 
the C=O carbonyl stretch (5.88\,$\mu$m), the C--H bending modes of
hydrocarbons (6.85 and 7.25\,$\mu$m), emission lines of H$_2$ S(5)
(6.91\,$\mu$m) and $[$Ar{\sc ii}$]$ (6.99\,$\mu$m), and methane ice
(7.67\,$\mu$m)}
\label{fig:5-8um}
\end{figure}

The 5.5--8\,$\mu$m spectral range of Galactic dense and diffuse ISM
lines of sight is rich in absorption features of vibrational transitions 
of various molecular species which are diagnostics of the chemical 
and physical states of these environments 
(\cite[Chiar et al. 2000; Keane et al. 2001]{chiar00,keane01}). 
Towards dense clouds, the profiles and strengths of the main volatile
ice absorption bands of water ice (6.0\,$\mu$m), NH$_4$$^+$ ice
(6.85\,$\mu$m) and methane ice (7.67\,$\mu$m) have been shown to vary 
considerably depending on the thermal history of the ice mantle. 
Observations of diffuse lines of sight, on the other hand, show 
absorption features in the 5.5--8\,$\mu$m range that are associated 
with organic refractory dust: e.g. the C--H bending modes of aliphatic
hydrocarbons at 6.85 and 7.25\,$\mu$m as seen towards the Galactic
Center (\cite[e.g. Chiar et al. 2000; 
Pendleton \& Allamandola 2002]{chiar00,pendleton02}). 
The strong detection by ISO of features of both the dense and the 
diffuse ISM towards several deeply enshrouded galactic nuclei
(\cite[Spoon et al. 2001, 2002]{spoon01,spoon02}) 
has expanded the applicability of these diagnostics to external
galaxies.

With the availability of IRS on Spitzer, both the wavelength 
coverage shortward of 5.7\,$\mu$m and the spectral resolution in the
5.5--8\,$\mu$m range have improved, enabling a better determination 
of the local continuum in this range, a clearer separation of the many 
spectral features and expansion of the study to a larger sample
of galactic nuclei.
Figure\,\ref{fig:5-8um} shows the optical depth spectra for five
deeply obscured merger nuclei. The top four spectra are dominated 
by strong absorption features centered at 6.0--6.1, 6.85 and
7.25\,$\mu$m. 
Following earlier studies (\cite[e.g. Chiar et al. 2000; 
Spoon et al. 2001, 2002, 2004b]{chiar00,spoon01,spoon02,spoon4b}), 
we identify the main components with water ice (6.0\,$\mu$m) and 
the bending modes of C--H in aliphatic hydrocarbons (6.85 and 
7.25\,$\mu$m).

Compared to the top four spectra, 
the optical depth spectrum of Arp\,220 (bottom panel) looks very
different. Strong PAH emission bands at 5.70, 6.22 and 7.7\,$\mu$m
and emission lines of H$_2$ S(5) and $[$Ar{\sc ii}$]$ at 6.91 and 
6.99\,$\mu$m distort the otherwise pure absorption spectrum. While
the two emission lines fill up the red wing of the 6.85\,$\mu$m 
hydrocarbon band, the blue flank of the 7.7\,$\mu$m PAH 
band overwhelms the 7.25\,$\mu$m hydrocarbon band and the 
6.22\,$\mu$m PAH emission feature fills up the depth of the 
6\,$\mu$m water ice band.
The same emission features are also weakly present in the 
spectrum of IRAS\,15250, most notably the PAH emission feature
at 6.22\,$\mu$m. Evidently, the contribution of unconcealed star
formation in IRAS\,15250 is far smaller than in Arp\,220.

At first glance, the spectral structure seen at 6.32\,$\mu$m in 
the spectra of IRAS\,08572 and IRAS\,00183 appears to be
the 6.22\,$\mu$m PAH feature. However, neither the central wavelength 
nor the FWHM of the feature is consistent with an identification 
with interstellar 6.22\,$\mu$m PAH. Instead, this feature may be 
`carved out' by absorption bands of species absorbing in this range, 
like gas phase water or hydrocarbons. For a detailed study of the 
absorption features contributing to the 6\,$\mu$m absorption complex 
we refer to Keane et al. (in preparation) and for a full analysis 
of the origin of the hydrocarbon bands in the spectrum of IRAS\,08572 
to Pendleton et al. (in preparation).

\section{Absorption features of C$_2$H$_2$, HCN and CO$_2$}

\begin{figure}
\begin{center}
\includegraphics[width=10.5cm]{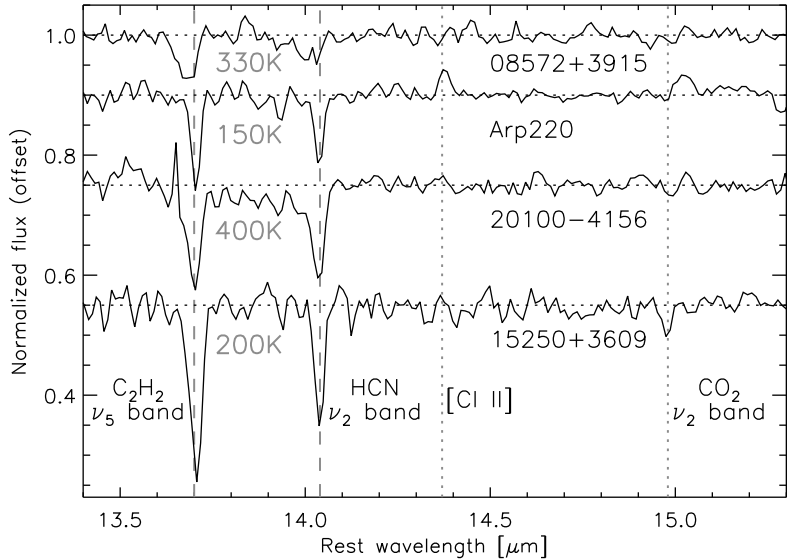}
\end{center}
\caption{Continuum-normalized high-resolution IRS spectra of four 
deeply obscured ULIRGs. The spectra reveal the presence of gas 
absorption bands of C$_2$H$_2$ at 13.7\,$\mu$m, HCN at 14.03\,$\mu$m 
and CO$_2$ at 15.0\,$\mu$m. Best fitting excitation temperatures 
for the C$_2$H$_2$ and HCN absorption bands are indicated on the plot.
Also indicated is the $[$Cl {\sc ii}$]$ line at 14.37\,$\mu$m in Arp\,220}
\label{fig:c2h2_hcn_profiles}
\end{figure}

Using the high-resolution mode of IRS on Spitzer, we have discovered
C$_2$H$_2$ (13.7\,$\mu$m), HCN (14.03\,$\mu$m) and CO$_2$ (15.0\,$\mu$m) 
gas absorption features in the spectra of a number of deeply obscured 
ULIRG nuclei (Armus et al., in preparation). Our four clearest detections
are shown in figure\,\ref{fig:c2h2_hcn_profiles}, together with
excitation temperatures resulting from a preliminary, combined fit 
of the C$_2$H$_2$ and HCN profiles. The excitation temperatures found 
range from 150\,K for Arp\,220 to 400\,K for IRAS\,20100.

These features have previously been seen in ISO--SWS spectra of deeply 
embedded massive Young Stellar Objects (YSOs) 
(\cite[e.g. Lahuis \& van Dishoeck 2000]{lahuis00}) and recently in 
the Spitzer spectra of a low mass YSO, IRS\,46 (Lahuis et al., in prep.). 
In the massive YSOs, the features originate in the hot inner dense region 
of the shells surrounding the young stars. In the case of IRS\,46 the 
features are believed to originate in the dense hot inner region of
the protostellar disk. In these cases, the chemistry is dominated by 
evaporation of the molecules from the grains with subsequent gas-phase 
processing which enhances the abundances of C$_2$H$_2$ and HCN by
orders of magnitude. Hence, C$_2$H$_2$ and HCN are typical tracers of 
high density ($>$ 10$^8$ cm$^{-3}$) high temperature chemistry.

The detection of absorption features of C$_2$H$_2$, HCN and CO$_2$ 
in our ULIRGs might suggest the presence of a population of deeply 
embedded YSOs in their nuclear environments. However, an identification 
with embedded protostars is not supported by spectral evidence 
in other parts of our mid-infrared spectra (e.g. by the absence of 
commonly detected NH$_4$$^+$ ice at 6.85\,$\mu$m and CO$_2$ ice 
at 15\,$\mu$m; \cite[Gibb et al. 2004]{Gibb04}). 
In the light of other evidence present (e.g. interferometric detection 
of HCN in these nuclei; extreme compactness of the ultra-luminous 
sources; weakness of common starburst emission features such as PAHs 
and fine-structure lines of Neon and Sulphur), it seems more likely 
that the features arise from highly pressure-confined star formation
in a dense medium --- i.e. under star formation conditions more
extreme than probed in ordinary starburst galaxies.
This, and a more detailed analysis of the 13--15\,$\mu$m features in 
ULIRG spectra, will be discussed in more detail in a future paper.

\section{The 4.65\,$\mu$m CO absorption feature}

\begin{figure}
\begin{center}
\includegraphics[width=10.5cm]{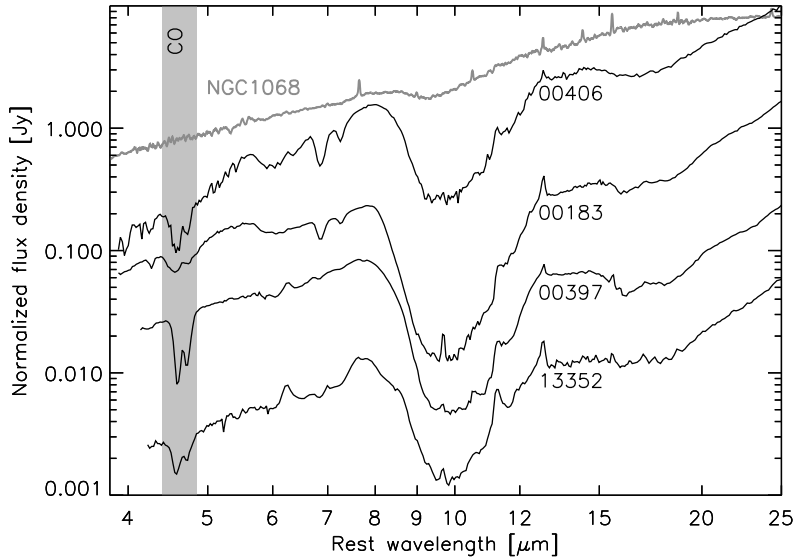}
\end{center}
\caption{4--25\,$\mu$m spectra of four ULIRGs showing an absorption
feature of gas phase CO at 4.65\,$\mu$m. The spectra are compared to
the ISO--SWS spectrum of the nucleus of the Seyfert-2 galaxy NGC\,1068
(\cite[Sturm et al. 2000]{sturm00}). The spectra have been scaled for 
plotting purposes}
\label{fig:co_seds}
\end{figure}

Absorption spectroscopy to study features of dust and gas in the 
2--5\,$\mu$m range requires the presence of a 2--5\,$\mu$m 
background continuum source. Active galaxies, with a favorable 
orientation of their AGN, show AGN-heated hot dust in their spectra
against which absorption features have been detected
(\cite[e.g. Imanishi et al. 2002]{imanishi02}). Also some of the 
deeply obscured galactic nuclei reveal 2--5\,$\mu$m absorption
features against their near-infrared dust continua 
(e.g. UGC\,5101: \cite[Imanishi et al. 2001]{imanishi01}; 
NGC\,4945: \cite[Spoon et al. 2000, 2003]{spoon00,spoon03}),
while for other nuclei these are impossible to detect given the 
absence of a hot dust background source (e.g. NGC\,4418: 
\cite[Spoon et al. 2001]{spoon01}; or IRAS\,15250: 
figure\,\ref{fig:sed_trends}). 

\begin{figure}
\begin{center}
\includegraphics[width=9cm]{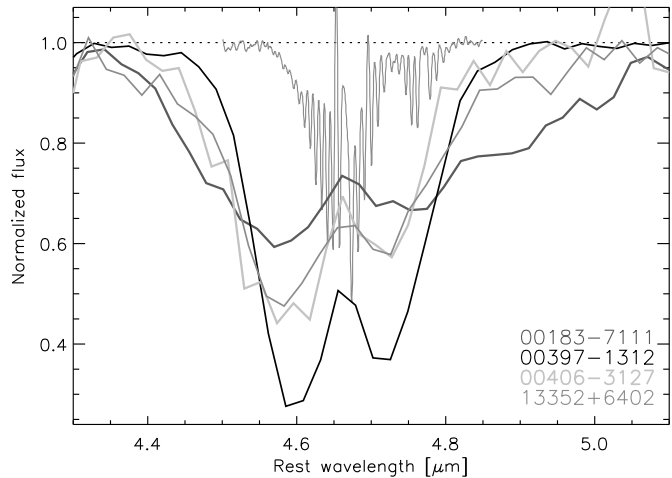}
\end{center}
\caption{Comparison of the normalized flux profiles for the CO 
absorption features in the spectra of Sgr\,A$^*$ 
(\cite[{\it dark grey} narrow profile; Moneti et al. 2001]{moneti01}) 
and the four ULIRGs shown in figure\,\ref{fig:co_seds}. Individual 
lines are resolved in the high-resolution ISO--SWS spectrum of Sgr\,A$^*$}
\label{fig:co_profiles}
\end{figure}

Here we report the first detection of strong absorption features
of 4.65\,$\mu$m gas phase CO in the spectra of four deeply obscured 
ULIRG nuclei with redshifts favorable to detection by Spitzer-IRS.
The 4--25\,$\mu$m spectra of the sources are compared
in figure\,\ref{fig:co_seds}, their normalized flux profiles are
shown in figure\,\ref{fig:co_profiles}. 

We have used the isothermal plane-parallel LTE gas models of Cami
(\cite[Cami 2002]{cami02}) to model the CO gas absorption profile
of IRAS\,00183. A best fit to the observed profile is found for a
model with a gas temperature of 720\,K, a column density of 
10$^{19.5}$\,cm$^{-2}$ and an intrinsic line width of 
50\,km/s. The temperature is far higher than found toward Galactic
ISM lines of sight, consistent with the clearly larger width of
the feature compared to for instance Sgr\,A$^*$ 
(figure\,\ref{fig:co_profiles}). Note that the density in the 
absorbing medium has to be high (n\,$>$\,3$\times$10$^6$\,cm$^{-3}$) 
to give rise to high J level absorption. This limits the size of the 
obscuring region to less than 0.03\,pc (\cite[Spoon et al. 2004b]{spoon04b}). 
Similar fits for the CO profiles of the other three sources in
figure\,\ref{fig:co_profiles} have not yet been made.

Following the discovery of hot CO gas in IRAS\,00183, \cite{lutz04} 
have analyzed a sample of ISO AGN spectra for the presence of similar 
CO absorption features and found none to good limits. This suggests 
that the CO features seen in our ULIRG spectra do not arise in a 
typical AGN torus geometry, but rather are a signature of the special
conditions in deeply obscured ULIRG nuclei --- pointing to a fully 
covered rather than a torus geometry (\cite[Lutz et al. 2004]{lutz04}).

\section{Conclusions}

Using the IRS spectrograph on the Spitzer Space Telescope, we have 
obtained mid-infrared spectra for a large sample of ULIRGs. The
spectra show a great diversity in spectral shape, reflecting the
diverse nature and merger evolutionary states of the sample. 

Especially interesting for astrochemists is the subsample of spectra 
showing signatures of strong obscuration, betraying the presence 
of huge amounts of dust and gas in and towards the merger nuclei.
The wide spectral coverage of the IRS (5--38\,$\mu$m), assisted by 
redshifts ranging from z=0.02 to z=0.4, allowed us to study the ISM
in these sources from rest frame $\sim$3.8\,$\mu$m to 37\,$\mu$m,
revealing absorption features of both amorphous and crystalline 
silicates, aliphatic hydrocarbons, water ice and gas phase bands 
of hot CO and warm C$_2$H$_2$, HCN and CO$_2$. PAH emission 
bands were found to be generally weak and in some cases absent.

Our analysis of the absorption features is far from complete and 
requires further comparison to the results from our emission line 
analysis to obtain a more detailed picture of the energetic
processes responsible for these rich and complicated spectra. 
It is clear, however, at this time that the features are probing 
a dense and warm environment in which crystalline silicates and 
water ice are able to survive but volatile ices, 
commonly detected in Galactic dense molecular clouds, cannot.
If powered largely by star formation, the stellar density and
conditions of the gas and dust have to be extreme not to 
give rise to the commonly detected emission features associated 
with starburst.

%
\begin{acknowledgments}
Support for this work was provided by NASA through the Spitzer Space
Telescope Fellowship Program, through a contract issued by the Jet
Propulsion Laboratory, California Institute of Technology under a 
contract with NASA
\end{acknowledgments}


\end{document}